\documentclass[a4paper]{jpconf}
\usepackage{graphicx,epsfig,color}
 \usepackage{amsfonts,amssymb}
\usepackage{amsmath}
\usepackage{youngtab}

\usepackage[% breaklinks=true,
colorlinks=true,backref,pagebackref]{hyperref}

\def\a{\alpha} \def\b{\beta} \def\g{\gamma} \def\d{\delta}
  
\def\vt{\vartheta} \def\vt{\vartheta} 
 \def\S{{\cal{S}}} \def\A{{\cal{A}}}

\def\a{\alpha}
\def\b{\beta}

\def\g{\gamma}
\def\d{\delta}
\def\g{\gamma}

\def\T{{\mathcal{T}}}

\def\C{{\mathcal{C}}}

\begin{document}
\title{Irreducible decompositions of the elasticity tensor under the linear and  orthogonal groups and their physical consequences}

\author{Yakov Itin$^1$ and Friedrich W.~Hehl$^2$}

\address{$^1$ Inst. Mathematics, Hebrew Univ. of Jerusalem \\
%\\E.J.\ Safra Campus,
%  Givat Ram,
%  Jerusalem 91904, Israel,\\
  and Jerusalem College of Technology, 
%21 Havaad Haleumi, POB 16031,\\
  Jerusalem 91160, Israel} \address{$^2$ Inst.\ Theor.\ Physics,
  Univ.\ of Cologne,
  50923 K\"oln, Germany\\
  and Dept.\ Physics \& Astron., Univ.\ of Missouri,\ Columbia, MO
  65211, USA} \ead{itin@math.huji.ac.il, hehl@thp.uni-koeln.de}

\begin{abstract}
  We study properties of the fourth rank elasticity tensor $C$ within
  {\it linear} elasticity theory.  First $C$ is irreducibly decomposed
  under the linear group into a ``Cauchy piece'' $S$ (with 15
  independent components) and a ``non-Cauchy piece'' $A$ (with 6
  independent components).  Subsequently, we turn to the physically
  relevant orthogonal group, thereby using the metric. We find the
  finer decomposition of $S$ into pieces with 9+5+1 and of $A$ into
  those with 5+1 independent components. Some reducible
  decompositions, discussed earlier by numerous authors, are shown to
  be inconsistent. --- Several physical consequences are
  discussed. The Cauchy relations are shown to correspond to
  $A=0$. Longitudinal and transverse sound waves are basically
  related by $S$ and $A$, respectively.  {\it file
    GhentCollGroups2014$\underline{\hspace{4pt}}$11.tex, 18 Nov 2014}

\end{abstract}

\section{Introduction}

The constitutive law in linear elasticity theory for an anisotropic
homogeneous body, the generalized Hooke law, postulates a linear
relation between the two second-rank tensor fields, the stress
$\sigma^{ij}$ and the strain $\varepsilon_{kl}$ (see
\cite{Love,SokolElast,Marsden,Hetnarski}):
\begin{equation}\label{hooke}
\sigma^{ij}=C^{ijkl}\,\varepsilon_{kl}\,.
\end{equation}
In standard linear elasticity, the stress and the strain tensors are
assumed to be symmetric. Consequently the elasticity tensor has the so
called {\it minor symmetries}\footnote{The Bach parenthesis $()$ and
  $[]$ denote symmetrization and antisymmetrization, respectively:
  $(ij):=\{ij+ji\}/2$ and $[ij]:=\{ij-ji\}/2$, see
  \cite{SchoutenRicci}.}
\begin{equation}\label{s1-sym}
C^{[ij]kl}= C^{ij[kl]}=0\,.
\end{equation}
Another restriction for the elasticity tensor is based on the energy
consideration.  The energy density of a deformed material is expressed
as $W={\scriptstyle\frac 12 }\,\sigma^{ij}\varepsilon_{ij}$. When the
Hooke law (\ref{s1-sym}) is substituted here, this expression takes
the form
\begin{equation}\label{energy1a}
W=\frac 12 \,C^{ijkl}\varepsilon_{ij}\varepsilon_{kl}\,.
\end{equation}
The right-hand side of (\ref{energy1a}) involves only those
combinations of the elasticity tensor components which are symmetric
under permutation of the first and the last pairs of indices.
Consequently, the so-called {\it major symmetry},
\begin{equation}\label{paircom}
  C^{ijkl}- C^{klij}=0\,
\end{equation}
is assumed. In 3-dimensional space, a fourth rank tensor with the
symmetries (\ref{s1-sym}) and (\ref{paircom}) has 21 independent
components.

In the literature on elasticity, a special decomposition of $C^{ijkl}$
into two tensorial parts is frequently used, see, for example,
Cowin \cite{Cowin1989}, Campanella \& Tonton \cite{Campanella},
Podio-Guidugli \cite{Podio-Guidugli}, Weiner \cite{Weiner}, and
Hauss\"uhl \cite{Haus}. It is obtained by symmetrization and
antisymmetrization of the elasticity tensor with respect to its {\it
  two middle indices:}
\begin{eqnarray}\label{MNdef}
M^{ijkl}:=C^{i(jk)l}\,,\quad
  N^{ijkl}:=C^{i[jk]l}\,,\qquad{\rm{with}}\qquad
C^{ijkl}=M^{ijkl}+ N^{ijkl}\,.
\end{eqnarray}

It is straightforward to show \cite{Itin-Hehl} that $M$ and $N$
fulfill the major symmetry (\ref{s1-sym}) but {\it not} the minor
symmetries (\ref{paircom}). Moreover, the tensor $M$ can be further
decomposed.  Accordingly, the reducible decomposition in
(\ref{MNdef})$_3$ does not correspond to a direct sum decomposition of
the vector space defined by $C$. The vector spaces of $M$ and $C$ both
turn out to be 21-dimensional, that of $N$ is 6-dimensional.  Thus,
the tensors $M$ and $N$ are auxiliary quantities but, due to the lack
of the minor symmetries, they they do {\it not} represent elasticity
tensors, that is, they cannot be used to characterize a certain
material elastically.  These quantities are placeholders without
direct physical interpretation. However, in the elasticity literature
some physical interpretations of the tensors $M$ and $N$ are
offered. Still, these results are inconsistent as we showed earlier in
\cite{Itin-Hehl}.

In this paper, we present various decompositions of the elasticity
tensor based on group-theoretic arguments and discuss some physical
applications of these decompositions.

The organization of the paper is as follows: In Sec.2 we discuss the
relation between the decomposition of a tensor and the group of
transformations acted on the basic vector space.  Sec.3 is devoted to
the $GL(3)$ decomposition of the elasticity tensor. It is based on the
Young diagram technique. Similarly, in Sec.4 we treat the case of the
$SL(3)$ and in Sec.5 that of the $SO(3)$. In Sec.6 we delineate some
physical applications of the irreducible decomposition described.
  
\section{Groups of transformations and corresponding 
decompositions of tensors}

In a {precise} algebraic treatment, a tensor must be viewed as a linear
map from a vector space to the field of real numbers, rather than as a
collection of real valued components.

A tensor of rank $p$ is defined as a multi-linear map from the
Cartesian product of $p$ copies of $V$ into the field ${\mathbb R}$,
\begin{equation}\label{tensor}
T:\underbrace{V\times \cdots\times V}_{p}\to \mathbb R\,.
\end{equation}
The set of all tensors $T$ of the rank $p$ compose a vector space by
itself, say ${\mathcal T}$. The dimension of this {\it tensor space}
is equal to $n^p$, that is, a 4th rank tensor in 3 dimensions (3d)
provides for ${\mathcal T}$ $3^4=81$ dimensions. As a basis in
${\mathcal T}$, we can take the tensor products of the basis elements
of $V$,
\begin{equation}\label{tensor-bas}
e_{i_1}\otimes e_{i_2}\otimes\cdots\otimes e_{i_p}\,.
\end{equation}
Accordingly, an arbitrary contravariant tensor of rank $p$ can be
expressed as
\begin{equation}\label{tensor-ex}
T=T^{{i_1} {i_2} \cdots {i_p}}\,e_{i_1}\otimes e_{i_2}\otimes\cdots\otimes e_{i_p}\,.
\end{equation}
Thus, the properties of the tensor $T$ in (\ref{tensor-ex}), in
particular its proper decomposition, is related to the group of
general linear transformations $GL(3)$ acting on the basis of $V$.
This group encompasses all non-singular $3\times 3$ matrices.

The vector space $V$ has a basis $e_i$ and a
  dual cobasis $\vt^i$. Under the $GL(3)$, the cobasis transforms as
  $\vt^{i'}=L_i{}^{i'}\vt^i$. We can attribute a volume element to
  this cobasis,
\begin{equation}\label{vol}
  \epsilon:=\frac{1}{3!}\epsilon_{ijk}\vt^i\wedge \vt^j\wedge \vt^k\,;
\end{equation}
here $\epsilon_{ijk}$ is the totally antisymmetric Levi-Civita
permutation symbol, which is $\epsilon_{ijk}=\pm 1$ for even or odd
permutations of $123$, respectively, and is $\epsilon_{ijk}=0$
otherwise. The primed volume element $\epsilon'$ for the transformed
cobasis $\vt^{i'}$ is defined analogously. If we substitute
$\vt^{i'}=L_i{}^{i'}\vt^i$, we find
\begin{equation}
  \epsilon'=\frac 1{3!}\epsilon
  _{i'j'k'}L_i{}^{i'}L_j{}^{j'}L_k{}^{k'}\vt^{i}\wedge \vt^{j}\wedge\vt^{k}\,.
\end{equation}
Since $\epsilon_{i'j'k'}=\epsilon_{ijk}=0,\pm 1$ is numerically
invariant under linear basis transformations, the volume element
transforms as a scalar {\it density,}
\begin{equation}\label{volumedensity}
\epsilon\,{'} = (\det L_{k}{}^{k'})^{-1}\,\epsilon\,.
\end{equation}

If we require that $\det L_{k}{}^{k'}\stackrel{!}{=}+1$, then the
volume element is an invariant (or scalar) and the $GL(3)$ reduces to
the special linear group $SL(3)$.

Furthermore, we know from elasticity theory that the description of a
deformation is based on the concept of the {\it distance} within a
continuum. A deformation is meant to be the {\it change} of the
distances between nearby points. However, the (Euclidean) distance
$ds^2=g:=g_{ij}dx^i\otimes dx^j$ as such, with $g_{ij}={\rm diag}
(1,1,1)$, is, by definition, invariant under the Euclidean group
$T(3)\rtimes SO(3)$, the semidirect product of the translations $T(3)$
with the rotations $SO(3)$. Accordingly, at one point in a continuum,
the special orthogonal group $SO(3)$ is the one relevant for the
representations of the tensors in elasticity theory.

The $SO(3)$ corresponds to all orthogonal matrices (reciprocal equals
to the transpose) with determinant $+1$. If we allow also parity
transformations, we arrive at the $O(3)$, the orthogonal group, and,
if we drop the orthogonality requirement, eventually at the
$GL(3)$. Alternatively, we can keep parity invariance in the first
step and then arrive at the $GL(3)$ via the special linear group
$SL(3)$, which has determinant $+1$:
\begin{align}\label{groups}
GL(3) \longleftarrow
\begin{Bmatrix}
  SL(3)\,, \det =+1\\
  O(3)\,,\hspace{5pt} \det =\pm 1
\end{Bmatrix}\longleftarrow SO(3)\,.
\end{align}

Incidentally, between the set of subgroups of $GL(3)$ we will use only
those that do not preserve any spatial direction. For physical
problems that involve some special direction, elastic material in
exterior magnetic field, for instance, smaller subgroups such as
$GL(2)$ are also relevant.

The process of going down from the $GL(3)$ to the $SO(3)$ is described
in detail in Schouten \cite{SchoutenPhysicist}, Chap.III.  On the
level of the $O(3)$, we can define a {\it scalar} volume
element\footnote{This is also discussed in some detail, for example,
  in \cite{Birkbook}, Secs.A.1.9 and C.2.3, respectively.} according
to $\eta:=\sqrt{\det( g_{kl})}\,\epsilon$.  The transformations of the
$O(3)$ are volume preserving. The $\epsilon$ is a {\it pre\/}metric
concept, whereas the $\eta$ requires the existence of a metric $g$.

We can collect there results in a table:
%\begin{align}
 % &(V,\epsilon) & \longrightarrow & & GL(3)\,,\nonumber\\
 % &(V,\epsilon\text{ with }\det L=+1) & \longrightarrow& &SL(3)\,,%\nonumber\\
%  &(V,\eta) & \longrightarrow& &O(3)\,,\nonumber\\
 % &(V,\eta\text{ with }\det L=+1)  & \longrightarrow& &SO(3)\,.%\nonumber
%\end{align}
\medskip

\begin{center}
    \begin{tabular}{| c | c | c | c |c|}
    \hline
    space & volume element & line element & transformation&group \\ \hline
    $V$ & $\epsilon$&  &  $\det L\ne 0$&$GL(3)$\\ \hline
   $V$ & $\epsilon$&  &  $\det L=+1$&$SL(3)$\\ \hline
   $V$ & $\eta$&$g$  &  $\det L=\pm 1 $&$O(3)$\\ \hline
$V$ & $\eta$&$g$  &  $\det L=+ 1 $&$SO(3)$\\ \hline
    \end{tabular}\medskip

{\it Table 1.} The different groups involved in the irreducible
decomposition of the elasticity tensor.
\end{center}

Although the group $SO(3)$ is highly relevant in elasticity theory, it
is convenient to provide the decomposition of the elasticity tensor
relative to the bigger groups and to arrive at the $SO(3)$ only at the
last stage. In this case, we are able to identify the different
origins of the irreducible parts and to derive the stratified
hierarchy of the invariants. Such a method is also useful from the
technical point of view because we need to use different algebraic
methods for different groups of transformations.  We begin with the
decomposition relative to the biggest group $GL(3)$.

\section{$GL(3)$-decomposition}\medskip

\subsection{Young decomposition}

Let us recall some basic facts about 4th rank tensors and their Young
decomposition:\vspace{-3pt}
\begin{itemize}

\item[1)] {Covariant tensor of 4th rank} over the vector space $V$
  (with ${\rm dim}V=3$) is a linear map from the Cartesian product of
  $4$ copies of the dual space $V^*$ into the real numbers,
\begin{equation}\label{map}
 {T:V^*\times V^*\times V^*\times V^*\to \mathbb R} \,. 
\end{equation}
\item[2)] The set of the 4th rank tensors constitutes a new vector
  space ${\cal T}$ called a {4th rank tensor space}. Its dimension is
  equal to $3^4=81$.

\item[3)] {The tensor space ${\cal T}$ can be decomposed into the
    direct sum of its subspaces that are invariant under the action of
    the group $GL(3,{\mathbb R})$. }

\item[4)] Due to the well-known {Schur-Weyl duality}, the irreducible
  decomposition of the space of the $p$th rank tensors under
  $GL(3,{\mathbb R})$ corresponds to the irreducible decomposition of
  the permutation group $S_4$.

\item[5)] A known practical way to derive the irreducible
  decomposition of $S_4$ is by use of {{Young diagrams. }}

\item[6)] A generic 4th rank tensor over the $3$d {tensor space can be
    decomposed into a direct sum of four subspaces.} This
  decomposition is depicted by the Young diagrams, see \cite{Boerner}
  or \cite{hamermesh},
\begin{equation}\label{irr3}
  \Yvcentermath1\yng(1)\otimes\yng(1)\otimes\yng(1)\otimes\yng(1)
  =\yng(4)\oplus \yng(3,1) \oplus \yng(2,2)\oplus \yng(2,1,1)\,.
  %\oplus \yng(1,1,1,1) \,.
\end{equation}
The left-hand side describes a generic 4th rank tensor.  On the
right-hand side, the diagrams represent the 4 subspaces.

\item[7)] The diagrams in (\ref{irr3}) come with a weight factor
  $f^\lambda$, called the {dimension of the irreducible representation}
  $\lambda$ of the permutation group $S_p$. For a diagram of the shape
  $\lambda$, the $f^\lambda$-factor is calculated by the use of the
  { hook-length formula},
\begin{equation}\label{hook1a}
f^\lambda=\frac{p!}{\prod_\a {\rm hook}(\a)}\,.
\end{equation}
Thus, for the diagrams depicted in (\ref{irr3}), 
\begin{equation}\label{hook1}
  f^{\lambda_1}=1\,,\quad f^{\lambda_2}=3\,,\quad\,f^{\lambda_3}=2\,,\quad
  f^{\lambda_4}=3\,,\quad f^{\lambda_5}=1\,.
\end{equation}

\item[8)] {The dimensions of the irreducible decomposition of
      $GL(n, \mathbb R)$} are calculated by the { Stanley
      hook-content formula}
    \begin{equation}\label{hook4} {\rm dim}\,
      V^\lambda=\prod_{(\a,\b)\in\lambda}\frac{n+\a-\b}{{\rm hook}(\a,\b)}\,.
\end{equation}
Consequently, for the diagrams depicted in (\ref{irr3}) and for the
dimension  $n=3$ of the vector space, we have
\begin{equation}\label{hook4a}
{\rm dim}\, V^{\lambda_1}=15,\quad{\rm dim}\, V^{\lambda_2}=15,\quad{\rm
  dim}\, V^{\lambda_3}=6,\quad{\rm dim}\, V^{\lambda_4}=3\,.
\end{equation}

%{\bf  (9)} 
\item[9)] Every diagram represents a subspace of the tensor space ${\cal T}$,
\begin{equation}\label{irr4} 
{\cal T}={}^{(1)}{\cal T}\oplus
{}^{(2)}{\cal T}\oplus {}^{(3)}{\cal T}\oplus {}^{(4)}{\cal T}\,.
\end{equation}
These subspaces intersect only at zero. Moreover, they all are
mutually non-isomorphic. Hence we have a { direct sum decomposition }
of the tensor space. The dimension of the initial tensor space is
separated into the dimension of the subspaces as follows:
\begin{equation}\label{hook5}
{\rm dim}\,{\cal T}= \sum_i f^{\lambda_i}\times{\rm dim}\,V^{\lambda_i}\,.
\end{equation}
The dimension of the tensor space is distributed between the subspaces
according to
\begin{equation}\label{hook5a}
 81=1\times 15+3\times 15+2\times 6 +3\times 3\,.
\end{equation}

\item[10)] The decomposition (\ref{hook5a}) is unique but
  reducible. In accordance with (\ref{hook1}), the subspaces
  ${}^{(2)}{\cal T}, {}^{(3)}{\cal T}$, and $ {}^{(4)}{\cal T}$ can be
  decomposed still further into sub-subspaces:
  \begin{equation}\label{irr4a}
    {}^{(2)}{\cal T}= \left({}^{(2,1)}{\cal
        T}\oplus{}^{(2,2)}{\cal T}\oplus{}^{(2,3)}{\cal T} \right)\,,
  \end{equation}
  \begin{equation}\label{irr4b}
    {}^{(3)}{\cal T}= \left({}^{(3,1)}{\cal T}\oplus{}^{(3,2)}{\cal T}\right)\,,
  \end{equation}
  \begin{equation}\label{irr4c} {}^{(4)}{\cal T}= \left({}^{(4,1)}{\cal
        T}\oplus{}^{(4,2)}{\cal
    T}\oplus  {}^{(4,3)}{\cal T} \right)\,.
\end{equation}
These decompositions are irreducible but not unique.

\end{itemize}

\subsection{Irreducible decomposition of $C^{ijkl}$}

The elasticity tensor is not a general 4th rank tensor; rather, it
carries its minor and major symmetries. Accordingly, we are looking
for a a decomposition of an invariant subspace of the tensor space
$\T$ into a direct sum of its sub-subspaces,
\begin{equation}\label{hook6x}
 \C=\a\,{}^{(1)}\!\C\,\oplus\,
\b\,{}^{(2)}\!\C\,\oplus\,\g\,{}^{(3)}\!\C\,\oplus\,\d\,{}^{(4)}\!\C\,,
\end{equation}
where
\begin{equation}\label{hook7}
\a=0,1\,;\qquad \b=0,1,2,3\,;\qquad \g=0,1,2\,;\qquad \d=0,1,2,3\,.
\end{equation}
Using the minor and major symmetries, we find $\a=1$ and $\d=0$.
Since ${\rm dim}\, \C=21$, we find as a unique solution of
(\ref{hook6x}),
$$\a=\g=1\,;\qquad \b=\d=0\,.$$
Thus, we proved \medskip

\noindent{{\sl Proposition 1:}} The decomposition
\begin{equation}\label{decomp}
  {C^{ijkl}=S^{ijkl}+A^{ijkl}}\,,\end{equation}
with 
\begin{equation}\label{firsty}
  S^{ijkl}:=C^{(ijkl)}=\frac 13(C^{ijkl}+C^{iklj}+ C^{iljk})\,,\quad\text{and}\quad
  A^{ijkl}:=\frac 13(2C^{ijkl}-C^{ilkj}-C^{iklj})\,,
\end{equation}
is irreducible and unique.\medskip

\noindent{{\sl Proposition 2:}} The partial tensors satisfy the minor
symmetries,
 \begin{equation}\label{alg1}
S^{[ij]kl}=S^{ij[kl]}=0\qquad {\rm and}\qquad  A^{[ij]kl}=A^{ij[kl]}=0\,,
\end{equation}
and the major symmetry,
 \begin{equation}\label{alg1a}
S^{ijkl}=S^{klij}\qquad {\rm and}\qquad  A^{ijkl}=A^{klij}\,.
\end{equation}

\noindent{{\sl Proposition 3:}} The irreducible decomposition of $C$
signifies the reduction of the tensor space ${\cal C}$ into a direct
sum of two subspaces ${\cal S}\subset{\cal C}$ (for the tensor $S$)
and ${\cal A}\subset{\cal C}$ (for the tensor $A$),
\begin{equation}\label{alg1b}
{{\cal C}={\cal S}\oplus{\cal A}\,.} 
\end{equation}
In particular, we have
\begin{equation}\label{alg1c} {{\rm dim} \, { C}=21\,,\qquad {\rm dim}
    \, {S}=15\,,\qquad {\rm dim} \, {A}=6\,.}
\end{equation}

\noindent{{\sl Proposition 4:}} The irreducible piece $A^{ijkl}$ of
the elasticity tensor is a fourth rank tensor. Alternatively, it can
be represented as a symmetric second rank tensor density
\begin{equation}\label{Del}
  \Delta_{mn}:=\frac 14\epsilon_{mil}  \epsilon_{njk}A^{ijkl}\,,
\end{equation}
where $\epsilon_{ijk}=0,\pm 1$ denotes the Levi-Civita permutation
symbol. The proof of this proposition is given in the next section.

\section{$SL(3)$-decomposition}

Since the elasticity tensor $C^{ijkl}$ satisfies the symmetries
(\ref{s1-sym}) and (\ref{paircom}), most of its contractions with
$\epsilon_{ijk}$ vanish.  Indeed, for the completely symmetric part
the contraction in two indices is identically zero,
\begin{equation}\label{S-contr}
\epsilon_{mij}S^{ijkl}=0\,.
\end{equation}

The second part of the elasticity tensor yields a non-vanishing
contraction,
\begin{equation}\label{S-contr-a}
  K_m{}^{jk}:=\frac 12 \epsilon_{mil}A^{ijkl}\,.
\end{equation}
This tensor $K_i{}^{jk}$ has vanishing traces and is antisymmetric
in the upper indices,
\begin{equation}\label{K-sym1}
  K_i{}^{ik}=0\,,\qquad K_k{}^{jk}=0\,,\qquad K_i{}^{(jk)}=0\,.
\end{equation}
Thus, $K_i{}^{jk}$ has 6 independent components, exactly like the
initial tensor $A^{ijkl}$.

Because of the antisymmetry of $K_i{}^{jk}$, we do not loose anything
if we contract the upper indices with $\epsilon_{njk}$:
\begin{equation}\label{Del-a}
  \Delta_{mn}:=\frac 12 \epsilon_{njk}K_m{}^{jk}=\frac
  14\epsilon_{mil}  \epsilon_{njk}A^{ijkl}\,.
\end{equation}
We can check that this tensor is symmetric
\begin{equation}\label{Del-sym}
\Delta_{[mn]}=0\,.
\end{equation}
Thus, it has the same 6 independent components. This proves the
Proposition 4. 

Summing up: there are no additional $SL(3)$ invariants of the
elasticity tensor, and this tensor cannot be decomposed further under
the action of the $SL(3)$. Moreover, we derived a representation of
$A^{ijkl}$ in terms of $\Delta_{mn}$. Under the action of the special
linear group $SL(3)$, the latter quantity is an ordinary tensor.
%
%Let me recall the well known stuff.  To state what is going with
%$\epsilon$ we need the low of its transformation.  We know it from the
%definition of the determinant. For a matrix $L^i{}_j$
%of %transformations between two basses $\vartheta^{i'}=L^{i'}{}_i\vartheta^i$,
%$$\det(L)=\frac 1{3!}\epsilon_{i'j'k'}\,\epsilon^{ijk}
%L_i{}^{i'}L_j{}^{j'}L_k{}^{k'}$$
%From here 
%$$\epsilon^{i'j'k'}=det(L^{-1})\epsilon^{ijk}L^{i'}{}_iL^{j'}{}_jL^{k'}{}_k$$
%

\section{$SO(3)$-decomposition}

Since for the elasticity tensor the invariance of the volume element
does not yield additional tensor decompositions, we skip the group
$O(3)$ and pass directly to its subgroup $SO(3)$.  Consider a vector
space $W$ endowed with a metric tensor $g_{ij}$. The norm and the
scalar product of vectors are defined now in the conventional way by $
(u,v)=g_{ij}u^iv^j$ and $ |v|=\left(v,v\right)^{1/2}$. In order to
preserve the scalar product, we must restrict ourselves to the
orthogonal group $O(3)$. Invariance of the volume element $\eta$ is
guaranteed if we restrict ourselves to the group $SO(3)$.  Now we can
use the metric tensor $g_{ij}$ and its inverse $g^{kl}$.

From the contraction of the metric with the totally symmetric Cauchy
part $S^{ijkl}$, a unique symmetric second rank tensor and its
scalar contraction can be constructed,
 \begin{equation}\label{S2}
   S^{ij}:=g_{kl}S^{ijkl}\qquad\text{and}\qquad S:=g_{ij}g_{kl}S^{ijkl}\,.
\end{equation}
Define the traceless part of the tensor $S^{ij}$ as
\begin{equation}\label{S00} {\not\!S}^{ij}:=S^{ij}-\frac 13
  Sg^{ij}\,,\qquad\text{with}\qquad g_{ij}{\not\!S}^{ij}=0\,.
\end{equation}
Now we turn to the decomposition of the tensor $S^{ijkl}$. We define
the subtensors
 \begin{equation}\label{sub1}
   {}^{(2)}\!S^{ijkl}:=\a {\not\! S}^{(ij}g^{kl)}\qquad\text{and}\qquad
   {}^{(3)}\!S^{ijkl}:=\b Sg^{(ij}g^{kl)}\,.
\end{equation}
We denote the remaining part as
\begin{equation}\label{sub3}
  {}^{(1)}\!S^{ijkl}:=S^{ijkl}-{}^{(2)}\!S^{ijkl}-{}^{(3)}\!S^{ijkl}\,.
\end{equation}
Now we require the tensor ${}^{(1)}S^{ijkl}$ to be traceless. This yields,
\begin{equation}\label{sub4}
 \a=\frac 67\,,\qquad\b=\frac 15\,.
\end{equation}
Hence we obtain the unique irreducible decomposition of the tensor
$S^{ijkl}$ into three pieces,
\begin{equation}\label{sub5}
S^{ijkl}={}^{(1)}\!S^{ijkl}+{}^{(2)}\!S^{ijkl}+{}^{(3)}\!S^{ijkl}\,.
\end{equation}
These pieces are invariant under the action of the group $SO(3)$.

In order to decompose the non-Cauchy part $A^{ijkl}$, it is convenient
to use its representation by the tensor density $\Delta_{ij}$. The
latter is irreducibly decomposed to a scalar and traceless parts.
\begin{equation}\label{sub6}
  \Delta_{ij}={\not\hspace{-4pt}\Delta}_{ij}+\frac 13\Delta
  g_{ij}\,,\qquad \text{where}\qquad \Delta:=g^{ij}
  \Delta_{ij}\,.
\end{equation}
Consequently, we obtain the decomposition of $A^{ijkl}$ into two
independent parts
\begin{equation}\label{sub7}
A^{ijkl}={}^{(1)}\!A^{ijkl}+{}^{(2)}\!A^{ijkl}\,,
\end{equation}
where {the scalar and the traceless parts are given by 
\begin{equation}\label{sub8}
{}^{(2)}\!A^{ijkl}:=\frac 23
\Delta\left(g^{ij}g^{kl}-g^{il}g^{jk}\right)\qquad\text{and}\qquad
{}^{(1)}\!A^{ijkl}:=A^{ijkl}-{}^{(2)}\!A^{ijkl} \,,
\end{equation}
respectively. This way we derived a composition of the elasticity
tensor into five irreducible parts
\begin{equation}\label{sub10}
{C^{ijkl}=\left({}^{(1)}\!S^{ijkl}+{}^{(2)}\!S^{ijkl}+{}^{(3)}\!S^{ijkl}\right)+
\left({}^{(1)}\!A^{ijkl}+{}^{(2)}\!A^{ijkl}\right)}\,.
\end{equation}
Between the first parentheses we collected the terms
corresponding to the Cauchy part, the second parentheses enclose the
non-Cauchy terms.  The dimension of the tensor space of the elasticity
tensor is decomposed into the sum of the corresponding dimensions of
the subspaces,
\begin{equation}\label{sub11}
21=\left(9+5+1\right)+
\left(5+1\right)\,.
\end{equation}
Thus, $C^{ijkl}\!$ can be represented by one totally traceless and totally
symmetric 4th rank tensor $^{(1)}\!C^{ijkl}\!$ {\it plus} two symmetric
traceless 2nd rank tensors ${\not\!S}^{ij},\,{\not\!\!\Delta}_{ij}$
{\it plus} two scalars $S,\,\Delta$.  This decomposition is unique and
irreducible under the action of the $SO(3,\mathbb R)$.

It is quite remarkable that the same decomposition was derived Backus
\cite{Backus} (see also Baerheim \cite{Baerheim}) in a rather
different way. In our group-theoretical treatment, we obtained all
irreducible pieces in a covariant form. Moreover, we derived a tree of
the independent invariances and their relations to different
transformation groups. We describe this stratified structure in the
following diagram:

%\vspace{20pt}
\begin{center}
\includegraphics[width=10truecm]{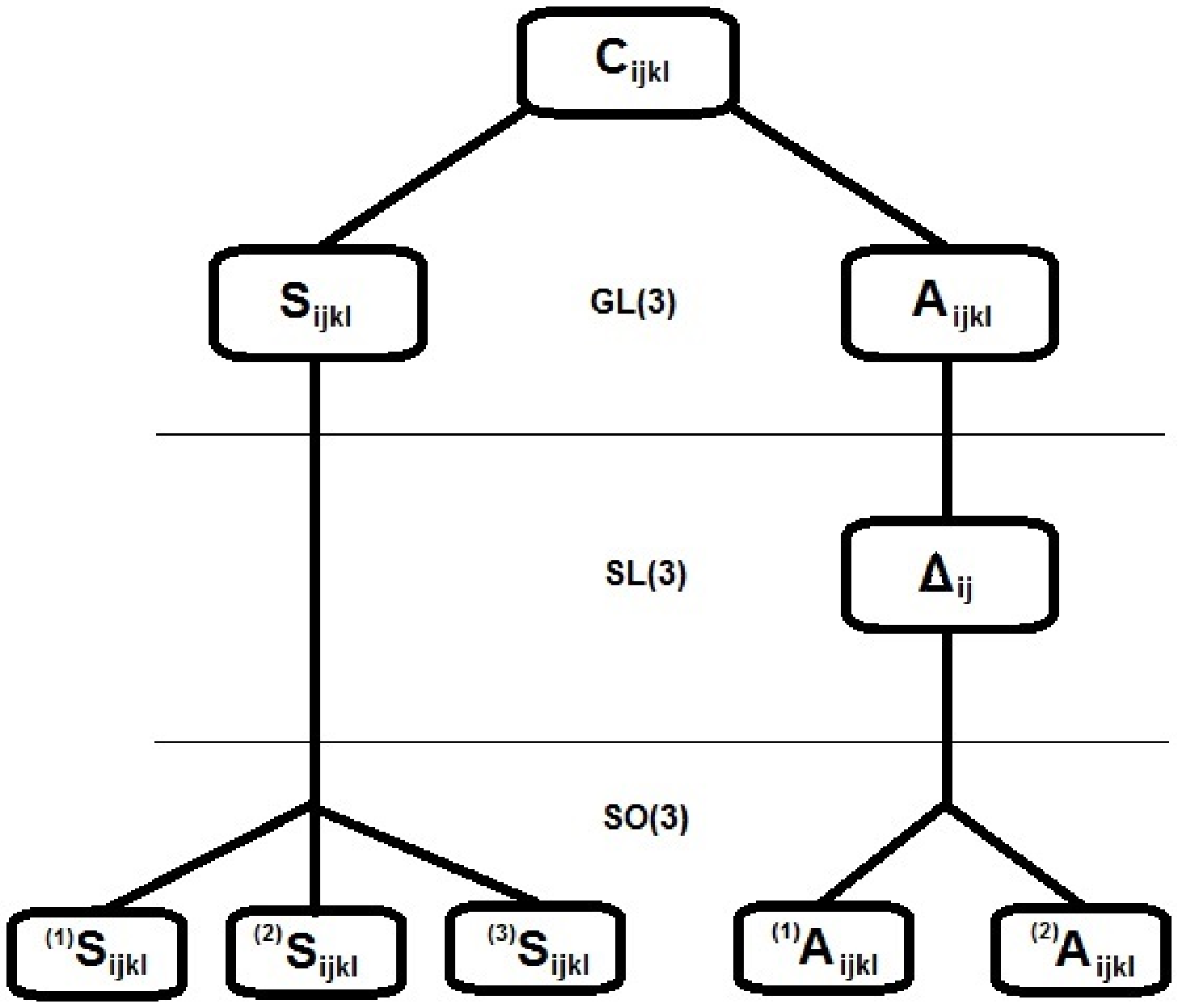}
\end{center}
\noindent {\it Figure 1.} The decomposition tree of the elasticity
tensor $C_{ijkl}$. Note that $\Delta_{ij}$ is a tensor (and {\it not}
a density) under the $SL(3)$.

\section{Applications} 

Some non-trivial applications were recently discussed by us in
\cite{Cauchy,Itin-Hehl} (for the analogous case in electrodynamics, see
\cite{Itin:2009aa}): 
\subsection{Cauchy relations}
The {Cauchy relations} are given by
\begin{equation}\label{cauchy1a}
  N^{ijkl}=0\qquad {\rm or} \qquad C^{ijkl}=C^{ikjl}\,,
\end{equation}
for their history, see Todhunter \cite{Todhunter}. The representation
in (\ref{cauchy1a}) is widely used in the elasticity literature, see, for
example,
\cite{Haus1983,Cowin1989,Cowin1992,Campanella,Podio-Guidugli,Weiner}.
In \cite{Cauchy,Itin-Hehl}, we presented an alternative form of these
conditions, namely
\begin{equation}\label{second}
  A^{ijkl}=0\qquad {\rm or} \qquad 2C^{ijkl}-C^{ilkj}-C^{iklj}=0\,.
\end{equation}
The last equation can be presented in a more economical form,
\begin{equation}\label{second-1}
\Delta_{mn}=0\,.
\end{equation}
The basic difference between (\ref{cauchy1a}) on the one hand and
(\ref{second}) or (\ref{second-1}) at the other hand is that our
conditions (\ref{second}) or (\ref{second-1}) are represented in an
irreducible form. This can have decisive consequences.

For most materials the Cauchy relations do not hold, even
approximately. In fact, the elasticity of a generic anisotropic
material is described by the whole set of the 21 independent
components, it is not restricted to the set of 15 independent
components obeying the Cauchy relations. This fact seems to nullify
the importance of the Cauchy relations for modern solid state theory
and leave them only as historical artifact.

However, a lattice-theoretical approach to the elastic constants
shows, see Leibfried \cite{Leibfried}, that the Cauchy relations are
valid provided (i) the interaction forces between the atoms or
molecules of a crystal are central forces, as, for instance, in rock
salt, (ii) each atom or molecule is a center of symmetry, and (iii)
the interaction forces between the building blocks of a crystal can be
well approximated by a harmonic potential{, see Perrin
  \cite{Perrin}}. Accordingly, a study of the {\it violations} of the
Cauchy relations yields important information about the intermolecular
forces of elastic bodies. Accordingly, one should look for the {\it
  deviation} $A^{ijkl}$ of the elasticity tensor $C^{ijkl}$ from its
Cauchy part $S^{ijkl}$.

\subsection{Acoustic waves}

If $u_i$ is the displacement field, the propagation of {acoustic
  waves} in anisotropic media is determined by the following
equation ($\rho$ = mass density): 
\begin{equation}\label{wave-eq} { \rho
    g^{il}\ddot{u_l}-C^{ijkl}\,{\partial_j\partial_k
      u_l}=0\,.}  \end{equation} With a plane wave
ansatz $%\begin{equation}\label{wave-an}
u_i=U_ie^{i\left(\zeta n_jx^j-\omega t\right)}$, %\,.  \end{equation}
we obtain a system of three homogeneous {algebraic}
equations 
\begin{equation}\label{char1} \left(
    v^2g^{il}-\Gamma^{il}\right)U_l=0\,, 
\end{equation} 
where the {\it Christoffel tensor} $\Gamma^{il}:=(1/\rho)\,
C^{ijkl}n_jn_k$ and of the {\it phase velocity} $v:=\omega/\zeta$ are
used.  Substituting the irreducible $S\!A$-decomposition into the
Christoffel tensor, we obtain 
\begin{equation}\label{Chr-dec1}
  \Gamma^{il}=\S^{il}+\A^{il}\,, 
\end{equation} 
where the Cauchy and non-Cauchy Christoffel tensors are
defined by
\begin{equation}\label{Chr-dec2}
  \S^{il}:=S^{ijkl}n_jn_k{=\S^{li}}\,,\qquad {\rm and }\qquad
  \A^{il}:=A^{ijkl}n_jn_k{= \A^{li} } \,.  
\end{equation} 

{Acoustic wave propagation} in an elastic {medium} is an eigenvector
problem, see (\ref{char1}), with the phase velocity $v^2$ as the
eigenvalues. In general, three distinct real positive solutions
correspond to three independent waves $^{(1)}U_l$, $^{(2)}U_l$, and
$^{(3)}U_l$, called {\it acoustic polarizations.}  For isotropic
materials, there are three pure polarizations: one {\it longitudinal
  (or compression) wave}} with $\vec{U}\times\vec{n}=0$ and two {\it
transverse (or shear) waves} with $ \vec{U}\cdot\vec{n}=0.$

In general, for anisotropic materials, three pure modes do not exist.
The identification of the pure modes and the condition for their
existence is an interesting problem.  Let us show how the irreducible
decomposition of the elasticity tensor, which we applied to the
Christoffel tensor, can be used here.

\noindent{{\sl Proposition 5:}}
Let $n^i$ denote {an allowed direction for the propagation of a {\it
    compression} wave.} Then the velocity $v_{\rm{L}}$ of { this wave}
in the direction of $n^i$ is determined only by the {\it Cauchy} part
of the elasticity tensor:
\begin{equation}\label{long-vel}
v_{\rm{L}}=\sqrt {S^{ij}n_in_j}\,.
\end{equation}
Moreover, for a {medium} with a given elasticity tensor, all three
{\it purely polarized} waves (one longitudinal and two {transverse})
can propagate in the direction {${\vec{n}}$} if and only if
\begin{equation}\label{long-cond}
S^{ij}n_j= S^{kl}n_kn_ln^i\,.
\end{equation}
Accordingly, for a given
medium, the directions of the purely 
polarized waves depend on the Cauchy part of the elasticity tensor
{\it alone}. In other words, two materials, with the same
Cauchy parts $S^{ijkl}$ {of the elasticity tensor} but
different non-Cauchy parts $A^{ijkl}$, have the same pure wave
propagation directions and the same longitudinal velocity.

%\section{Conclusions}
%\section*
\noindent {\bf Acknowledgments.} This work was supported by the
German-Israeli Foundation for Scientific Research and Development
(GIF), Research Grant No.\ 1078-107.14/2009. Y.I.\ would like to thank
Jan-Willem Van Holten for the invitation to the Group 30 Colloquium.
%----------------------------------------------------


\medskip
\medskip
\begin{thebibliography}{99}

\bibitem{Backus} Backus, G. 1970 Geometrical picture of anisotropic
  elastic tensors. {\it Reviews Geophys.\ Space Phys.} {\it 8},
  633--671.

\bibitem{Baerheim} Baerheim, R. 1993 Harmonic decomposition of the
  anisotropic elasticity tensor. {\it Quarterly J.\ Mech.\ Appl.\
    Math.} {\it 46}, 391--418.

\bibitem{Boerner} Boerner, H. 1970 {\it Representations of
    groups}. Amsterdam: North Holland.

\bibitem{Campanella} Campanella, A.\ \& Tonon, M.~L. 1994 A note on the
  Cauchy relations, {\it Meccanica} {\it 29}, 105--108.

\bibitem{Cowin1989} Cowin, S.~C. 1989 Properties of the anisotropic
  elasticity tensor. {\it Quarterly J.\ Mech.\ Appl.\ Math.} {\it
    42}, 249--266. Corrigenda 1993 ibid.\ {\it 46}, 541--542.

\bibitem{Cowin1992} Cowin, S.~C.\ \& Mehrabadi, M.~M. 1992 The structure of
  the linear anisotropic elastic symmetries. {\it J.\ Mech.\ Phys.\
    Solids} {\it 40}, 1459--1471.

\bibitem{hamermesh} Hamermesh, M. 1989 {\it Group Theory and its
    Application to Physical Problems.} New York: Dover.

\bibitem{Haus1983} Hauss\"uhl, S. 1983 {\it Physics of Crystals (in
    German).} Weinheim, Germany: Physik-Verlag.

\bibitem{Haus} Hauss\"uhl, S. 2007 {\it Physical Properties of
    Crystals: An Introduction.} Weinheim, Germany: Wiley-VCH.

\bibitem{Cauchy} Hehl, F.~W.\ \& Itin, Y. 2002 The Cauchy relations in
  linear elasticity theory.  {\it J.\ Elasticity} {\it 66}, 185--192
  [\href{http://arXiv.org/pdf/cond-mat/0206175}{arXiv:cond-mat/0206175}].
% arXiv.org:cond-mat/0206175

\bibitem{Birkbook} Hehl, F.~W. \ \& Obukhov, Yu.~N. 2003 {\it
    Foundations of Classical Electrodynamics,} Charge, flux, and
  metric. Boston, MA: Birkh\"auser.

\bibitem{Hetnarski} Hetnarski, R.~B.\ \& Ignaczak, J. 2011 {\it The
    Mathematical Theory of Elasticity,} 2nd edn. Boca Raton, FL: CRC
  Press.

\bibitem{Itin:2009aa} Itin, Y. 2009 On light propagation in premetric
  electrodynamics. Covariant dispersion relation, {\it J.\ Phys.\ A}
  {\it 42}, 475--402.
   
\bibitem{Itin-Hehl} Itin, Y. \& Hehl, F.~W. 2012 The constitutive
  tensor of linear elasticity: its decompositions, Cauchy relations,
  null Lagrangians, and wave propagation, {\it J.\ Math.\ Phys.\ 54},
  042903 (2013)
  [\href{http://arXiv.org/pdf/1208.1041}{arXiv:1208.1041}].
% arXiv:1208.1041

\bibitem{Leibfried} Leibfried, G. 1955 Gittertheorie der mechanischen
  und thermischen Eigenschaften der Kristalle. In {\it Handbuch der
    Physik/Encyclopedia of Physics (ed.\ S.~Fl\"ugge), vol.\ VII/1,
    Kristallphysik I,} pp.\ 104--324. Berlin: Springer.

\bibitem{Love} Love, A.~E.~H. 1927 {\it A Treatise on the Mathematical
    Theory of Elasticity,} 4th edn. Cambridge, UK: University Press.

\bibitem{Marsden} Marsden, J.~E.\ \& Hughes, T.~J.~R. 1983 {\it
    Mathematical Foundations of Elasticity.} Englewood Cliffs, NJ:
  Prentice-Hall.

\bibitem{Perrin} Perrin, B. 1979 Cauchy relations revisited, {\it
    Phys.\ Stat.\ Sol.\ B} {\it 91}, K115--K120.

\bibitem{Podio-Guidugli} Podio-Guidugli, P. 2000 {\it A Primer in
    Elasticity.} Dordrecht: Kluwer.

\bibitem{SchoutenRicci} Schouten, J.~A. 1954 {\it Ricci-Calculus}, 2nd
  edn. Berlin: Springer.

\bibitem{SchoutenPhysicist} Schouten, J.~A. 1989 {\it Tensor Analysis
    for Physicists,} reprinted 2nd edn. New York: Dover.

%\bibitem{Sokol} Sokolnikoff, I.~S. 1951 {\it Tensor Analysis.} New
%  York: Wiley.
%
\bibitem{SokolElast} Sokolnikoff, I.~S. 1956 {\it Mathematical Theory
    of Elasticity,} 2nd edn. New York: McGraw-Hill.

\bibitem{Todhunter} Todhunter, I. 1960 {\it A History of} {\sl the
    Theory of Elasticity} {\it and the Strength of Materials, from
    Galilei to Lord Kelvin,\/} edited and completed by K.\ Pearson.
  Vol.\ I: Galilei to Saint-Venant 1639--1850, pp.496--505. New York:
  Dover [orig.\ publ.\ in 1886].

\bibitem{Weiner} Weiner, J.~H. 2002 {\it Statistical Mechanics of
    Elasticity.}  Mineola, NY: Dover.

\end{thebibliography}
\end{document}